\documentclass[11pt]{article}
\usepackage[a4paper,margin=2.5cm]{geometry}
\usepackage{graphicx}
\usepackage{amsmath,amssymb}
\usepackage{bm}
\usepackage{xcolor}
\usepackage{setspace}
\usepackage[super,sort&compress,comma]{natbib}
\usepackage{url}

\usepackage[colorlinks=true,allcolors=blue]{hyperref}
\hypersetup{pdftitle={Collective order from language models without communication, memory or shared context}}

\title{Collective order from language models without communication, memory or shared context}
\author{Yong Wang\\[2pt]
\small Department of Physics and Materials, The Hong Kong Polytechnic University,\\ \small Hung Hom, Kowloon, Hong Kong, 999077, China\\[2pt]
\small \texttt{physics-yong.wang@polyu.edu.hk}}
\date{}

\begin{document}
\maketitle

\noindent\textbf{%
Physical matter orders without a centre: units interact only with neighbours, and fluctuations carry them over energy barriers from a locally stable arrangement to the ordered state.
We ask whether language-model agents can do the same, treating each memoryless query as a particle that sees its neighbours but cannot communicate, has no identifier or shared context and receives the same question as all others.
Such a particle behaves as if at zero temperature: told a rule or a temperature, it returns the most probable state, and a collective can freeze in a local minimum.
We therefore give each particle only a goal and a die whose weights it sets itself, and test several models.
Some of them use the die as a noise source even at a local optimum, with weights that fall with the energy cost of the move; on a lattice the energy of one of them climbs over barriers and reaches the ground state, where the noise stops by itself.
Particles in three-lane counterflow also order, from mixed traffic into one-way lanes: a spontaneous symmetry breaking.
Decentralized collective order can thus arise from disorder without communication, memory or shared context, with noise that the particles choose themselves.}

\bigskip

Order without a centre is common in nature.
Magnetic moments order below a critical temperature, and self-propelled particles that only align with their neighbours under noise move as a flock~\cite{Vicsek1995,TonerTu1995,TonerTu1998}; such active systems form a field of their own~\cite{Marchetti2013,VicsekZafeiris2012}.
Animals do the same: starlings respond to a fixed number of nearest neighbours yet are correlated across the whole flock~\cite{Ballerini2008,Cavagna2010,Bialek2012}, marching locusts switch from disorder to order as their density rises~\cite{Buhl2006}, and a few informed individuals can steer a group without signals or recognized leaders~\cite{Couzin2005,Sumpter2006}.
In each case identical units follow one local rule and the whole acquires an order that no unit holds~\cite{Anderson1972}.
Fluctuations decide whether order appears.
In frustrated or glassy systems they do more: they let a unit occasionally take a state that is not the best for it, so that a region trapped in a locally optimal arrangement can escape, the effect that simulated annealing turns into a general method of optimization~\cite{Kirkpatrick1983}.
In condensed matter, too, thermal fluctuations let particles move together: simulations show atoms rearranging cooperatively in a supercooled liquid, in superionic conductors and in a compressed crystal~\cite{Donati1998,He2017superionic,Wang2021calcium}.
Without fluctuations two obstacles remain that a centre would otherwise remove: identical units acting deterministically and at the same time on identical observations cannot break their symmetry~\cite{Itai1990symmetry,Fich2003impossibility}, and units that each take their own best option can freeze in local optima of the collective task.

Collectives of large-language-model (LLM) agents have been organized in many ways, and the tools of statistical physics have begun to be applied to them~\cite{Song2025detailed}.
Decentralized designs with shared ledgers, task queues or self-chosen roles have been studied~\cite{Ding2025dmas,Mao2026delm,Dochkina2026hierarchy}, as have agents on rings, cycles and networks~\cite{Hirota2026local,Parsaee2025loopbench,Grotschla2025agentsnet} and systems for open-ended search with a program that evaluates, selects or validates the candidates~\cite{RomeraParedes2024,Zhang2025dgm,Ren2026elite}.
Whether LLM agents can order as matter does depends, as in matter, on whether they have a source of fluctuation.
Language models sample poorly from a distribution stated in words~\cite{Xiao2025flipping,Zhao2026dice,Gu2026illusion}, and raising the sampling temperature does not repair this~\cite{Zhao2026dice,Gu2026illusion}: it changes the wording of answers but not how often a costlier option is chosen (Supplementary Section~S8).
Noise supplied from outside has been used to let agents sample~\cite{Gu2026illusion,Ezaki2026shared,Mahmud2025dsa}, and players of a zero-sum game have been given a uniform random suggestion that they could follow or ignore~\cite{Yang2025coin}.
Identical, memoryless agents on a square lattice, each seeing its four neighbours, have been studied without a task, with the sampling temperature as the control parameter~\cite{DeNobili2026lattice}.
In the work we are aware of, it remains open whether memoryless agents that see only their neighbours, with no messages, identifiers, shared context or noise supplied by a designer, can order, and, where their own answers do not vary, whether a die given without any rule for its use is rolled where it is needed.

We ask this in its barest form, treating each agent as a particle: it has no memory, receives no messages and carries no identifier or shared context, and every particle receives the same query, which shows it only its own state and those of its neighbours.
Because a particle sees only its neighbourhood, its response can be measured one particle at a time before any collective is built, and the collective can be compared with what that response predicts.
We do not prescribe the noise: we give the particle a goal and a die, a tool that rolls one of its options with weights given in the answer, and leave to the model whether to roll, with which weights, and whether to follow the outcome.
We compare four models from four vendors, DeepSeek V4.1 Flash, Qwen3.8-27B, GPT-6.2 Astra and Claude Opus 5.5.

The answer comes in four steps.
A rule or a temperature stated in words does not work and leaves the answer of every model at the best option, and a random number handed in from outside only lets the model compute a rule its designer already knew.
Given a goal and a die, one of the four models tested systematically leaves a local optimum on a roll of the die at the cost of an extra conflict, the second obstacle named above, and its single-particle rates anticipate which models' lattices order.
On a lattice of $81$ particles the two runs of that model agree with the prediction: the number of conflicts rises over barriers, falls to zero, and the rolling stops.
Particles that move against each other in lanes can sort out of mixed traffic into one-way lanes, a spontaneous breaking of symmetry; the die is not needed for this, but it lets the traffic of a model whose answers repeat flow much more smoothly.
In each step where the die is used, whether to roll and with what weights is left to the particle.

\begin{figure}[tbp]
\centering
\includegraphics[width=\textwidth]{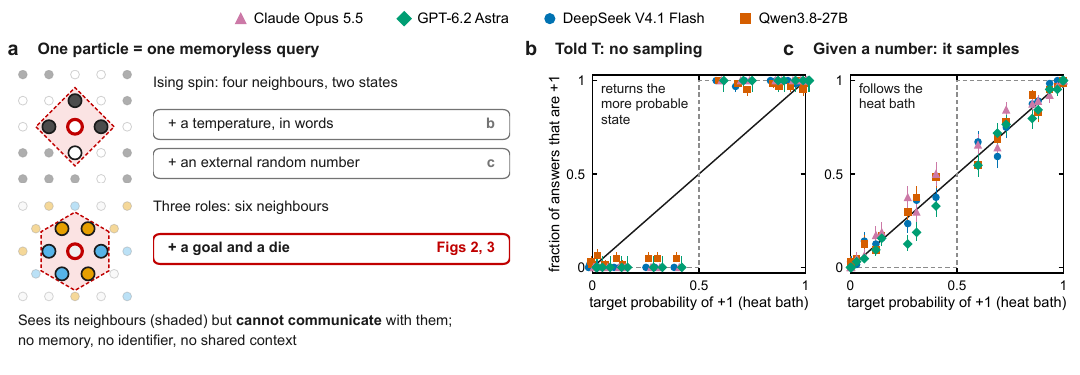}
\caption{\label{fig:one}\textbf{Language-model particles, and why words do not supply their fluctuation.}
\textbf{a}, Every particle receives the same query, sent anew at each update, showing only its own state and its neighbours' (shaded); it cannot communicate and has no memory, identifier or shared context. Top, an Ising spin (four neighbours, two states), used in \textbf{b} and \textbf{c}; bottom, a particle with three roles on the triangular lattice (Figs~\ref{fig:two} and \ref{fig:three}).
\textbf{b}, An Ising spin told its temperature $T$ under heat-bath dynamics: fraction of answers $+1$ against the heat-bath probability $p_+=1/(1+e^{-2h/T})$ (diagonal), where $h$ is the sum of the four neighbouring spins; dashed, the more probable state.
\textbf{c}, The same with a uniform random number from an external program. $T$ and $h$ in Methods.}
\end{figure}

\subsection*{Noise handed in from outside}
Every particle receives the same query, which shows it only its own state and those of its neighbours [Fig.~\ref{fig:one}a]. A preference of a model for a particular name or position cannot account for the order reported here: in every collective task the names of the options and the order in which neighbours are listed are drawn at random in every query (Methods and Supplementary Section~S14). We first measured a single spin of the two-dimensional Ising model.
Given the scoring rule, every model adopts the majority of its neighbours in $256$ of $256$ queries: under a stated rule the agent always takes the better state, as a particle at zero temperature would (Supplementary Fig.~S1).
Words do not change this.
Told that it is a spin at temperature $T$ under heat-bath dynamics~\cite{Glauber1963}, the rule under which a spin takes each state with its Boltzmann probability, every model returns the more probable state at every temperature, although DeepSeek V4.1 Flash and Qwen3.8-27B state the correct Boltzmann probability in their reasoning [Fig.~\ref{fig:one}b]; asked to choose at random between two equally good states, the models give label-dependent or deterministic answers rather than a coin (Supplementary Fig.~S1).
Many replies, or the reasoning returned with them, give the reason: the heat-bath probability is computed correctly, but no random number is available to use it; in the words of one reply, ``I have no random number generator here, so I took the more likely outcome rather than drawing a true random sample'' (Supplementary Section~S7).
In each of these tests one answer is best under the stated rule, and there the answers of a model do not vary; whether they vary where several moves are nearly equally good is taken up with the moving particles below.

The fluctuation can be supplied from outside.
Given a uniform random number from an external program, returned by a tool, the agents compare it with the heat-bath probability and their choices fall on the heat-bath curve [Fig.~\ref{fig:one}c].

But a particle that receives its noise this way adds nothing that its designer did not already specify.
The form of the noise, its amount and the places where it acts were all fixed by us, a centralized way, and the model evaluates a rule known in advance at the cost of one query per update, where a line of code would do.
Where randomness has been brought to language-model agents, it has been as a random number or a stochastic search loop supplied from outside~\cite{Gu2026illusion,Ezaki2026shared,Mahmud2025dsa}.
We therefore stop specifying the noise.

\begin{figure}[tbp]
\centering
\includegraphics[width=\textwidth]{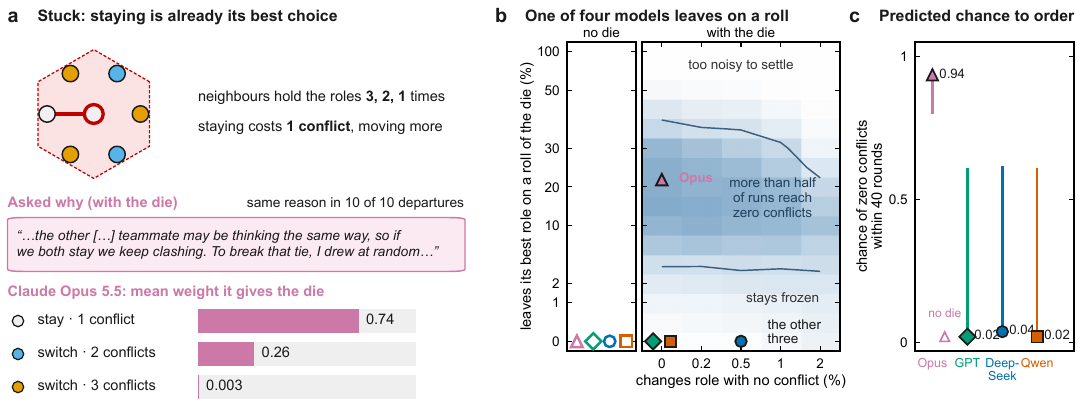}
\caption{\label{fig:two}\textbf{Single particles given a goal and a die.}
\textbf{a}, The local optimum: the neighbours hold the three roles $3$, $2$ and $1$ times and the particle the rarest, so staying costs $1$ conflict and switching $2$ or $3$ (red bond, its conflict). Bars, mean die weights of Claude Opus 5.5 in $96$ queries (a query without a roll counts as staying). Box, part of one reply when asked for a reason, with $[\ldots]$ for the role letter (Methods; full replies in Supplementary Section~S12).
\textbf{b}, Each model by its two single-particle rates: role changes without a conflict and departures from the local optimum on a weighted roll of the die (Methods). Filled, with the die; open, left strip, without it. Shading, fraction of runs of the reference rule that reach zero conflicts within $40$ rounds; lines, half of the runs; both axes are stretched near zero.
\textbf{c}, Probability of zero conflicts within $40$ rounds that the reference rule gives at each model's rates; lines, range over the $95\%$ interval of the departure rate; open, Claude Opus 5.5 without the die (Supplementary Table~S12).}
\end{figure}

\subsection*{Letting the model roll the die}
We give the agents only a goal and a die, and leave the rest to the model's answer: whether to roll, with which weights on its options, and whether to follow the outcome.
We state no temperature and no rule, and nothing tells the agent where randomness belongs; the die is a source of noise that the model may use or ignore (see Methods for detailed prompts).
The task is one on which greedy local updates get stuck.
Every particle on a triangular lattice takes one of three roles, and each of its six neighbours with the same role counts as a conflict; the number of conflicts plays the role of an energy, that of the antiferromagnetic three-state Potts model, so that a state without conflicts has the lowest energy~\cite{Wu1982,Schick1977,Adler1995}.
A particle with one neighbour in its own role and two and three in the others already has the fewest conflicts, yet still has one [Fig.~\ref{fig:two}a].
At such a local optimum no move lowers the energy, and greedy particles couldn't reach a conflict-free state.
The prompt states the goal, that nobody should have a conflict, and offers the die (Methods).

Measured one particle at a time, before the full lattice was run, the four models differ in one rate.
Almost no particle changes its role without a conflict.
At the local optimum, Claude Opus 5.5 leaves on a weighted roll of the die in $21$ of $96$ queries, and the other three models almost never do; without the die no model leaves [Fig.~\ref{fig:two}b; Methods].
Its mean weights for staying, for one extra conflict and for two are $0.74$, $0.26$ and $0.003$ [Fig.~\ref{fig:two}a]: it mostly stays, sometimes steps one conflict up and almost never takes the worst role.
Read as a Boltzmann factor, the ratio $0.74{:}0.26$ is an effective temperature $T\approx1$ in units of one conflict, but no single temperature fits all three weights (Supplementary Section~S3).
Asked for its reason in a separate set of $96$ queries (Methods), Claude Opus 5.5 gives the same one in all $10$ replies that leave: the neighbour holding the same role sees the same situation, so if both stay the conflict remains, and a roll weighted towards staying breaks the tie at the cost of one more conflict [Fig.~\ref{fig:two}a].
This is the symmetry obstacle named in the introduction. On the lattice, neighbours never update at the same time, so what the roll does there is simpler: it lets the particle leave the local optimum.
This behaviour is not unique to one model: in a few informal trials of the same query, Gemini 4 Neon also proposed a roll with weights that fall with the number of conflicts and gave the same reason; we did not test it systematically, and the prompt is given in Methods for readers who wish to try other models.

These two rates anticipate the lattice.
A reference rule based on this single-particle behavior (Methods) gives Claude Opus 5.5 a probability of $0.94$ of reaching zero conflicts within $40$ rounds, and the other three models at most $0.08$ [Fig.~\ref{fig:two}c].

\begin{figure}[tbp]
\centering
\includegraphics[width=\textwidth]{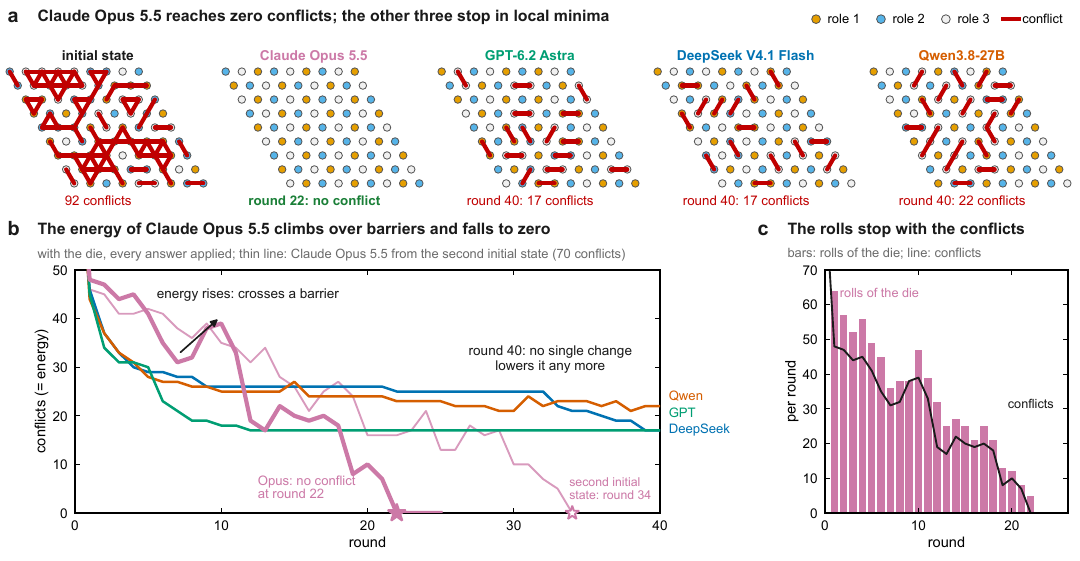}
\caption{\label{fig:three}\textbf{The lattice test.}
$81$ particles on a $9\times9$ periodic triangular lattice with the goal-and-die prompt; four models, the same initial state and update order, every answer applied. Red bonds mark conflicts, not messages.
\textbf{a}, The initial state and each model's state at the end (for Claude Opus 5.5, its first conflict-free state).
\textbf{b}, Number of conflicts (the energy) versus round; stars, conflict-free state; thin line, Claude Opus 5.5 from a second initial state.
\textbf{c}, Claude Opus 5.5: particles that rolled the die in each round (bars) and conflicts (line).}
\end{figure}

\subsection*{Crossing barriers on the lattice}
We ran $81$ particles of each model on a $9\times9$ periodic lattice from the same random initial state and update order in several independent rounds [Fig.~\ref{fig:three}].
In the language of the Potts model, a state without conflicts is the ground state; a configuration in which no single change lowers the number of conflicts is a local minimum, and leaving it requires a temporary rise, a barrier.
GPT-6.2 Astra, DeepSeek V4.1 Flash and Qwen3.8-27B descend quickly and at round $40$ are in local minima, with $17$, $17$ and $22$ conflicts [Fig.~\ref{fig:three}a,b].
Claude Opus 5.5 leaves local optima on weighted rolls in about a quarter of such updates, and almost every move that raised a particle's conflicts was such a roll (Methods; Supplementary Section~S3).
Its energy therefore climbs over barriers and falls to the ground state, at round $22$ and, from the second initial state, at round $34$ [Fig.~\ref{fig:three}b].
There the rolls stop [Fig.~\ref{fig:three}c]: it never rolled without a conflict (Supplementary Section~S3).

\begin{figure}[tbp]
\centering
\includegraphics[width=\textwidth]{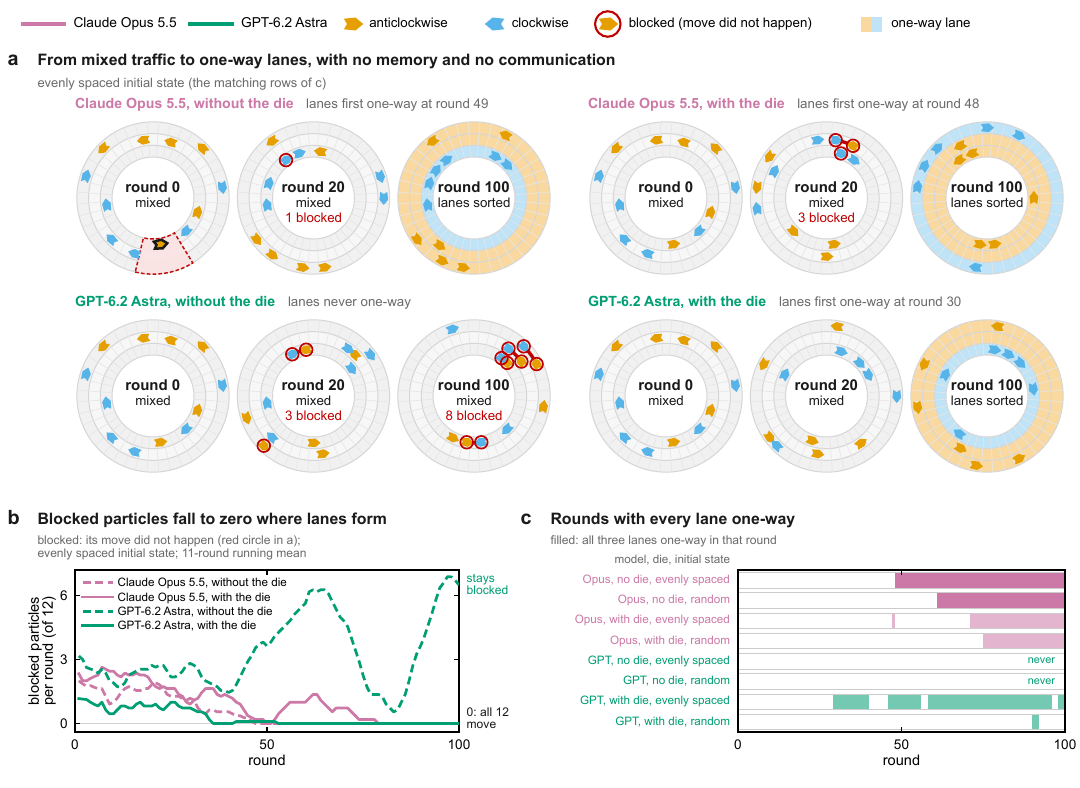}
\caption{\label{fig:flock}\textbf{From disorder to collective lanes.}
Twelve particles on a periodic track of $40$ cells with three lanes, drawn as three concentric rings, six moving each way; each sees the three lanes two cells ahead and behind (pink sector), with no memory and no communication.
Two initial states: evenly spaced, one particle in each of $12$ equal segments of the ring with two of each direction in every lane; and random, $12$ cells drawn at random with several conflicts (Methods).
\textbf{a}, Snapshots at rounds $0$, $20$ and $100$ from the evenly spaced initial state (random, Supplementary Fig.~S8). Red circle, a blocked particle, whose move did not happen in that round; tinted lanes, every lane one-way. Different runs settle into different assignments of directions to lanes.
\textbf{b}, Blocked particles per round ($11$-round running mean), evenly spaced initial state; the counterpart of the conflicts in Fig.~\ref{fig:three}b.
\textbf{c}, Rounds in which all three lanes are one-way (filled), for all eight runs.
Space-time plots, Supplementary Fig.~S7.}
\end{figure}

\begin{figure}[tbp]
\centering
\includegraphics[width=\textwidth]{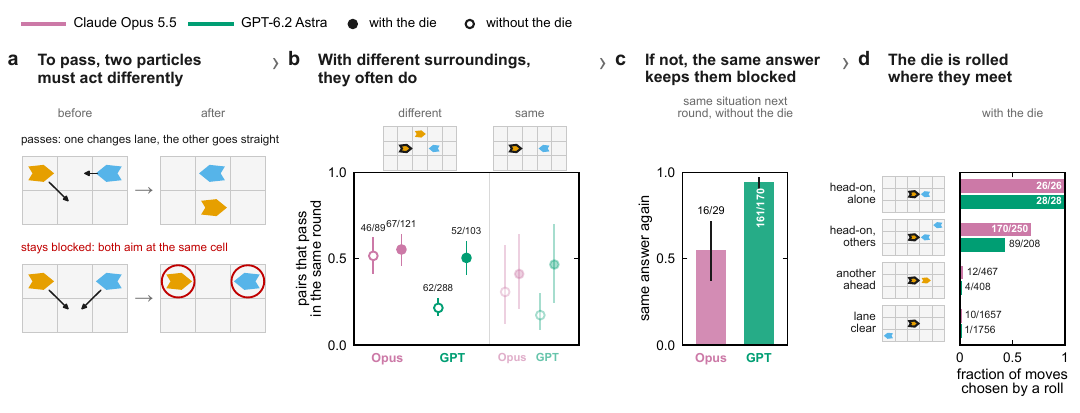}
\caption{\label{fig:five}\textbf{Why lanes can form without a die, and where the die is rolled.}
Runs of Fig.~\ref{fig:flock}, both initial states pooled; small grids show what a particle (bold) sees; the panels follow one argument from \textbf{a} to \textbf{d}.
\textbf{a}, Two particles meeting head-on pass only if they move differently; the same move leaves them blocked (red circles).
\textbf{b}, Pairs that pass in the same round (at least one of the two moves), when the two see different surroundings or exactly the same cells (the latter few, shown faint); filled, with the die; open, without; bars, $95\%$ Wilson intervals.
\textbf{c}, Without the die, how often a blocked particle that meets the same situation in the next round gives the same answer.
\textbf{d}, With the die, fraction of moves chosen by a roll in each situation.}
\end{figure}

\subsection*{From disorder to collective lanes}
Collective behavior is one of the most widespread forms of order in nature. Insect bands, bird flocks and human crowds organize their movement although each member responds only to its neighbours, with no leader and no global plan~\cite{Buhl2006,Ballerini2008,Helbing1995}, and similar minimal models describe collective motion from cells and bacteria to crowds~\cite{Vicsek1995,VicsekZafeiris2012,Marchetti2013}. It is therefore the natural test for particles that cannot communicate: whether their local choices add up to coherent motion.
Twelve particles travel along three lanes side by side, six in each direction, on a track of $40$ cells with periodic boundaries, so that a particle leaving one end re-enters at the other; Fig.~\ref{fig:flock} therefore draws the track as a ring, with the lanes as three concentric circles; each sees the three lanes two cells ahead and behind, and moves one cell ahead, one cell ahead into a neighbouring lane, or stays [Fig.~\ref{fig:flock}a] (see Methods for detailed prompts).
A move into an occupied cell, two moves into the same cell and two moves that cross do not happen and count as a clash, and the shared goal is that every particle travels as smoothly as it can with as few clashes as possible.
The flow is ordered when every lane carries a single direction, as in the lanes of counterflowing pedestrians and driven colloids~\cite{Helbing1995,Dzubiella2002}, where strong noise instead jams the flow~\cite{Helbing2000freezing}.
We ran Claude Opus 5.5 and GPT-6.2 Astra with and without the die, from an evenly spaced and a random initial state, for $100$ rounds; the prompts of the two arms differ by the one sentence that offers the die, and animations of the other runs are provided with the code (Methods).

In these runs, particles with no memory and no communication sort out of mixed traffic into one-way lanes [Fig.~\ref{fig:flock}a].
From the evenly spaced initial state lanes form in three of the four conditions, for Claude Opus 5.5 with and without the die and for GPT-6.2 Astra with the die; from the random state both runs of Claude Opus 5.5 sort [Fig.~\ref{fig:flock}c; Supplementary Fig.~S8].
In the last $20$ rounds of the five runs that end sorted, every particle moves in every round.
The sorting is a spontaneous breaking of symmetry: nothing sets which lane carries which direction, and the five sorted runs settle into different assignments, the inner lane carrying the anticlockwise direction in two and the clockwise direction in three [Fig.~\ref{fig:flock}a,c].
The lattice, too, has several equivalent ground states; in both settings these are observations on small systems.

The die is not necessarily needed for lanes to form, but it makes the traffic of GPT-6.2 Astra much smoother: from either initial state, without it about three of the twelve particles cannot move in an average round and the lanes never sort, with it fewer than one and a half [Fig.~\ref{fig:flock}b; Supplementary Table~S2].
The particles that cannot move play the role of the conflicts on the lattice [compare Fig.~\ref{fig:three}b].

Lanes can form without a die because most encounters are not deadlocks, situations that only a random choice could resolve [Fig.~\ref{fig:five}].
Two particles meet head-on when they face each other in the same lane, adjacent or one cell apart, and they pass only if their moves do not clash: for example one changes lane while the other goes straight, whereas if both aim at the same cell, as when both change into the same cell of the middle lane, neither moves [Fig.~\ref{fig:five}a].
In most encounters the two particles see different surroundings, for example a third particle in view of only one of them, so the same prompt shows them different situations and their answers often differ: without the die Claude Opus 5.5 passed in about half of such encounters within the round, GPT-6.2 Astra in about a fifth [Fig.~\ref{fig:five}b].
A blocked particle whose surroundings do not change meets the same situation in the next round, and only a different answer can then free it: in such repeats the answer of Claude Opus 5.5 changed about every other time, that of GPT-6.2 Astra about one time in twenty [Fig.~\ref{fig:five}c; answers to one query asked $24$ times are in Supplementary Table~S4].
For answers that repeat, this is where the die acts.
With it, GPT-6.2 Astra passed in about half of the encounters with different surroundings, as often as Claude Opus 5.5, and both models rolled in about half or more of the head-on updates and in almost none where the lane ahead was clear [Fig.~\ref{fig:five}d].
For Claude Opus 5.5, whose answers already vary, the die acts in these runs as additional noise, and the system relaxes more slowly: its lanes settled for good later than without the die, at round $76$ instead of $62$ from the random initial state and at round $72$ instead of $49$ from the evenly spaced one, where they first sorted at round $48$ and then mixed again, much as the energy on the lattice occasionally rises on its way down [Figs~\ref{fig:flock}b,c and \ref{fig:three}b].
Where a barrier separates the particles from order, as on the lattice, the noise is needed; where encounters are not deadlocks, it only slows the relaxation.

\subsection*{Discussion}
Order without a centre needs two ingredients, local interactions and fluctuations.
A language-model particle brings the first but not the second: told a rule or a temperature it returns the best state, like a particle at zero temperature, and a collective of such particles can freeze in a local minimum.
In the language of statistical physics, the number of conflicts can be seen as an energy, and the die is a local source of noise, a heat bath only in the loose sense that it supplies uphill moves.
At a local optimum one answer is best, and without the die no model leaves it; Claude Opus 5.5 rolls there, with mean weights $0.74$, $0.26$ and $0.003$ for staying, one extra conflict and two extra conflicts, based on $96$ single-particle queries at the local optimum [Fig.~\ref{fig:two}a], and never where it has no conflict.
Seen this way, Claude Opus 5.5 switches the noise on where the energy is locally but not globally minimal, gives less weight to moves that cost more, and switches the noise off in the ground state: each particle anneals itself, locally, without a cooling schedule set from outside.
A heat bath cannot do this.
At a fixed temperature a system either stays below the barriers or, once it has crossed them, is shaken out of the ground state again; noise that follows the local state lets the system cross the barriers and still come to rest, because the ground state becomes absorbing and detailed balance is broken, the structure that focused local search and decentralized colouring build in by hand~\cite{Selman1994noise,Seitz2005focused,Duffy2008decentralised}.

In matter the fluctuations are supplied by the environment; here the particle chooses them, and the reason Claude Opus 5.5 states for its rolls is itself a symmetry argument: identical particles in identical situations would make identical choices, and only randomness can tell them apart [Fig.~\ref{fig:two}a].
The lanes show the other side of symmetry.
Nothing in the rules sets which lane carries which direction, and different runs settled into different assignments, a spontaneous breaking of symmetry.
Where several moves are nearly degenerate, the answers themselves can fluctuate enough to let two particles pass; where a gap separates the best move, as at a local optimum, they freeze.
Fluctuation at near-degeneracy and freezing across a gap tie the two settings together: a language-model particle needs noise from outside only where its own answers freeze.

Treating agents as particles lets one ask of them the questions asked of matter: measure the response of one particle, predict the collective, and read off energy, barriers and broken symmetry.
Our systems are small, each collective condition was run several rounds, and the particles are four specific model versions, so the results are observations rather than statistics and say nothing yet about phase transitions.
The next steps are larger systems, a map of how the order depends on how often a particle leaves a local optimum and how often it moves without need, and other tasks.
For now each update is a model call of seconds, so these collectives evolve slowly, but latency is a matter of engineering, not of principle.
The particles have three properties that suit large decentralized systems: without messages the cost grows with the number of particles, not of links; without memory or identifiers any unit can be replaced, added or lost with no state to synchronize; and a stated goal replaces a hand-tuned temperature or annealing schedule.
If an update took milliseconds, such particles could allocate shared resources and coordinate their motion in real time, with no dispatcher and from local observation alone.
Whether this holds at large sizes and in real time depends on the scaling and the additional runs listed above.
What the experiments show is a particle whose microscopic rule includes the choice of its own noise; the statistical physics of such particles has just begun.

\bibliographystyle{naturemag}
\bibliography{refs}

\section*{Methods}

\subsection*{Particles and queries}
Every update of every particle is one independent request to a model, with no conversation history.
The message states the particle's own current state, the current states of its neighbours, the rule or goal, and the answer format; the particle is not told where it sits, how large the system is, which round it is, or what anyone answered before. The detailed prompts are in English, and are shown in the following.
In every query the program chooses anew which letter denotes which state and in which order the neighbours are listed, so that a fixed preference for a letter or a position becomes, in terms of the physical state, an unbiased tie-break.

\subsection*{Models and access}
We used four reasoning models: DeepSeek V4.1 Flash (\texttt{deepseek-v4.1-flash-1017}), Qwen3.8-27B (\texttt{qwen3.8-27b}), GPT-6.2 Astra (\texttt{gpt-6.2-astra}) and Claude Opus 5.5 (\texttt{claude-opus-5-5}). All queries were made between 16 and 23 November 2026. Gemini 4 Neon was tried informally on the local-optimum query a few times and is not among the four models tested systematically.
DeepSeek V4.1 Flash and Qwen3.8-27B were queried with a requested sampling temperature of $1.0$, reasoning effort ``low'' and a limit of $20\,000$ output tokens; their reasoning text is returned and stored but never used to determine the answer.
Claude Opus 5.5 and GPT-6.2 Astra use their default reasoning effort, and the reasoning is not returned. To obtain the replies shown in Fig.~\ref{fig:two}a, the local-optimum situation was posed to Claude Opus 5.5 with the explicit prompt, its last sentence replaced by a request to state first, in one or two sentences, what the particle sees and why it chooses as it does. This test is used only for the reasons; all rates in the text are from the tests without the request for all models.

\subsection*{Randomized labels}
In the three-role tests the assignment of the letters A, B and C to the roles (six possibilities) and the order in which the six teammates are listed change from query to query; in the Ising tests with letters the assignment of A and B and the order of the neighbours do the same. In the three-lane runs the names of the lanes and the order of the listed cells are drawn anew in every query from $36$ combinations; the Ising temperature tests use the fixed labels $+1$ and $-1$ and identical wording within each cell. What is randomized in each experiment is listed in Supplementary Section~S14.
The die draws from the weights given by the agent with a random number of the program.
We checked this randomization separately, by reconstructing every query from its seed and with $2936$ additional queries to DeepSeek V4.1 Flash.
The assignments and orders are uniform and independent of each other and of the situation, the die samples the given weights correctly, and the $31\,033$ recorded random numbers of the die are uniform and without serial correlation.
A model may prefer some letters to others; with the assignments balanced, such a preference cancels, and weighting the six assignments equally changes each main readout by at most $0.1$ percentage points.
The code that generates the prompts and the die will be released with the next version of this manuscript.

\subsection*{Single-particle tests, Ising setting}
A spin with four neighbours is shown its own state and those of its neighbours as the letters A and B, with a scoring rule under which the state shared with more neighbours scores better, without that rule, or with an added sentence for ties (Supplementary Section~S7).
In the temperature tests the spins are shown as $+1$ and $-1$ and the agent is told, without a formula, that it is a spin of the two-dimensional Ising model at temperature $T$ under heat-bath dynamics; $T=0.5$, $1.5$, $2.27$, $4$ and $10$, with $32$ identical queries for each $T$, local field $h$ (the sum of the neighbour spins) and own state.
With the random number the prompt is word for word the same; the tool, which returns a uniform number in $[0,1)$, is offered outside the text.
A decision with the tool is consistent if $s=+1$ exactly when the number is below $p_+(h,T)=1/(1+e^{-2h/T})$.

\subsection*{Three-role problem}
Each particle on a periodic triangular lattice takes one of three roles; each of its six neighbours with the same role counts as one conflict, and the total number of conflicts is the energy of the antiferromagnetic three-state Potts model.
The prompt with the goal and the die reads as follows (an example of a particle at a local optimum; letters and neighbour order are random in every query):
\begin{quote}\small\ttfamily
You are one member of a team. Each member takes one of three roles: A, B or C.\\
Your role right now is: C\\
You work directly with six teammates. Their roles right now are:\\
First: A\\
Second: B\\
Third: A\\
Fourth: C\\
Fifth: A\\
Sixth: B\\
Rule: each teammate who has the same role as you counts as one conflict.
The team's goal is that nobody has any conflict.\\
Each member sees only their own six teammates.\\
\end{quote}\vspace{-\medskipamount}\begin{quote}\small
\noindent\fcolorbox{red!70!black}{red!10}{\parbox{\dimexpr\linewidth-2\fboxsep-2\fboxrule\relax}{\ttfamily A tool named draw is available: given a weight for each of A, B and C, it returns one role drawn at random with probability proportional to the weights. You may use it or not.}}
\end{quote}\vspace{-\medskipamount}\begin{quote}\small\ttfamily
Now choose your role.\\
No need to explain. End your final reply with a line that contains only
"Answer:" followed by A, B or C.
\end{quote}
The prompts with and without the die differ only in the highlighted sentence and are otherwise word for word the same.
A particle's situation is given by how many neighbours hold each role and which is its own, $16$ situations in all.
The test with the die has $440$ queries, $96$ of them at the local optimum (neighbours $3$--$2$--$1$, own role on the $1$) and $200$ in conflict-free situations with a unique best role.
For GPT-6.2 Astra, Qwen3.8-27B and DeepSeek V4.1 Flash the $96$ local-optimum queries were supplemented by $400$, $2000$ and $2000$ further queries with the same prompt.
A control poses the local optimum with the same prompt without the sentence on the tool and without any tool ($96$ queries for Claude Opus 5.5 and GPT-6.2 Astra, $400$ for Qwen3.8-27B and DeepSeek V4.1 Flash).
Two rates summarize a response table: how often a particle without a conflict changes its role, and how often a particle at the local optimum leaves it.

\subsection*{Lattice runs}
The lattice has $9\times9$ sites with periodic boundaries; since 9 is a multiple of 3, the conflict-free states, in which each of the three sublattices holds one role, fit the lattice.
Each round updates the three sublattices one after another in a random order, each in parallel.
All models start from the same two random initial states ($92$ and $70$ conflicts), with the same sequence of orders, for $40$ rounds or until three rounds after the first conflict-free state; every answer is applied.
A departure from a local optimum counts as made on a weighted roll of the die when the answer is the role the die returned and the roll gave positive weight to staying and to at least one other role, with weights that do not increase with the number of conflicts.
Over rounds $1$--$40$ and both initial states, Claude Opus 5.5 left a local optimum on a weighted roll of the die in $364$ of $1373$ updates, GPT-6.2 Astra in $0$ of $2124$, DeepSeek V4.1 Flash in $0$ of $2564$ and Qwen3.8-27B in $4$ of $2425$.
By the same criterion, single particles at the local optimum left in $21$ of $96$ queries for Claude Opus 5.5, $0$ of $496$ for GPT-6.2 Astra, $0$ of $2096$ for Qwen3.8-27B and $4$ of $2096$ for DeepSeek V4.1 Flash; without the die no model left ($0$ of $96$ for Claude Opus 5.5 and GPT-6.2 Astra, $0$ of $400$ for the other two).

\subsection*{Counterflow in three lanes}
The track has $40$ cells in each of three parallel lanes, with periodic boundaries; a cell holds at most one particle, and six particles move each way.
Each particle sees the cells two ahead and two behind in all three lanes ($14$ cells), each empty or holding a particle that faces the same or the opposite way, and moves one cell ahead in its own lane, one cell ahead into a neighbouring lane, or stays.
All particles move at once; a move into an occupied cell, two moves into the same cell and two moves that would cross do not happen, and each counts as one clash for the particle that tried it.
In the evenly spaced initial state one particle sits in each of $12$ segments of three or four cells, and every lane holds two particles of each direction; in the random initial state $12$ of the $120$ cells are drawn at random.
Each of the eight arms (two models, with and without the die, two initial states) is one run of $100$ rounds.
The lanes are sorted when no lane holds particles moving both ways, and a particle is blocked in a round when its move does not happen.
Two particles in the same lane that face each other, adjacent or one cell apart, form a head-on encounter; the pair passes in a round if at least one of the two moves.
The prompt calls the particles birds and their direction of motion a wind. The example below is for a particle in an outer lane. One other particle faces it in the same lane, with one empty cell between them, and no other particle is in view. With the die, the prompt reads:
\begin{quote}\small\ttfamily
You are a bird on a track of square cells with three lanes, A, B and C, that run side by side: lane B lies between lane A and lane C. A cell holds at most one bird.\\
Cells are named by lane and by offset along the track from your own cell, which is A 0. Offset +1 is the next cell on the side you are facing, offset -1 is the next cell on the opposite side, and so on.\\
A bird can stay where it is or move one cell toward the side it faces, ending in its own lane or in a lane next to its own. Your moves are named by the lane they end in:\\
A: to A +1\\
B: to B +1\\
The wind you feel right now blows toward the + side.\\
A bird faces the way the wind it feels blows. Each bird sees only the cells around itself and chooses its own move.\\
You can see the cells with offsets from -2 to +2 in all three lanes. Right now:\\
A +2: a bird facing the - side\\
A +1: empty\\
A -1: empty\\
A -2: empty\\
B +2: empty\\
B +1: empty\\
B 0: empty\\
B -1: empty\\
B -2: empty\\
C +2: empty\\
C +1: empty\\
C 0: empty\\
C -1: empty\\
C -2: empty\\
A move into a cell that is occupied does not happen. If two birds try to move into the same cell, neither of them moves. If the moves of two birds would cross each other, neither of them moves.\\
Rule: a move that does not happen counts as one clash for the bird that tried it. The birds' shared goal is that every bird travels as smoothly as it can along the wind it feels, with as few clashes as possible.\\
\end{quote}\vspace{-\medskipamount}\begin{quote}\small
\noindent\fcolorbox{red!70!black}{red!10}{\parbox{\dimexpr\linewidth-2\fboxsep-2\fboxrule\relax}{\ttfamily A tool named draw is available: give it a weight for any of A, B and Stay (an option you leave out gets weight 0), and it returns one of them drawn at random with probability proportional to the weights. You may use it or not.}}
\end{quote}\vspace{-\medskipamount}\begin{quote}\small\ttfamily
Now choose your move.\\
No need to explain. End your final reply with a line that contains only ``Answer:'' followed by A, B or Stay.
\end{quote}
The prompts with and without the die differ only in the highlighted sentence and are otherwise word for word the same.
The code that builds, randomizes and sends the queries of all experiments and analyses the replies is described under Code availability, together with animations of the complete lane runs.

\subsection*{Reference lattice simulations}
Reference simulations use the same lattice and update order, with no model call, from random initial states ($400$ runs per point).
Let $n_X$ be the number of neighbours of a particle in role $X$ and $s$ its own role, so that $n_s$ is its number of conflicts and the energy is $E=\tfrac12\sum_i n_{s_i}$.
At each update the particle takes the role
\begin{equation}
s' =
\begin{cases}
\text{one of the other two roles at random, with probability } p_c, & n_s = 0,\\[2pt]
\arg\min_{X\neq s} n_X \text{ (at random if tied), with probability } p_h, & 0 < n_s < n_X \text{ for all } X\neq s,\\[2pt]
\text{a role with the fewest conflicts, at random among ties}, & \text{otherwise},
\end{cases}
\end{equation}
and keeps $s$ when the first or the second move is not made.
The first line is a change without need, the second a departure from a local optimum, and the third a downhill or level move.
The success map $S(p_c,p_h)$ (shading in Fig.~\ref{fig:two}b) is the fraction of runs that reach $E=0$ within $40$ rounds.
The predictions of Fig.~\ref{fig:two}c are $S$ at the rates $(p_c,p_h)$ measured for each model on single particles; their ranges follow from the $95\%$ Wilson interval of $p_h$, which gives $S$ from $0.80$ to $0.96$ for Claude Opus 5.5 and at most $0.14$ for the other models.
The rule is a reference with two parameters, not a real simulation of the agents.

\subsection*{Statistics and limitations}
Counts are given with their denominators; intervals are Wilson intervals, $68\%$ in Fig.~\ref{fig:one} and $95\%$ elsewhere.
The main conclusions do not rest on single trajectories. The single-particle rates are counted over hundreds to thousands of independent queries, with option names and listing order drawn at random, and the differences between models are far larger than their intervals. The lattice runs agree with the outcome that the reference simulation anticipates from these rates; the lattices of Claude Opus 5.5 and GPT-6.2 Astra and every condition of the lanes were run from independent initial states, with the reported outcome in both. Larger systems, more initial states and other wordings of the prompt remain to be tested.

\subsection*{Data availability}
All data used in this paper will be deposited in a public repository upon publication. Please email the author if you need them in advance; the author is happy to share them and discuss.

\subsection*{Code availability}
The code that builds, randomizes and sends the queries of all experiments and analyses the replies, together with every prompt word for word, is available at \url{https://github.com/yongwangxx/Decentralized-Collective-Multi-Agents} and archived at \url{https://doi.org/10.5281/zenodo.24461514}. Animations of the complete lane runs are at \url{https://yongwangxx.github.io/Decentralized-Collective-Multi-Agents/}.

\subsection*{Use of AI tools}
Large language models were used as tools to assist in writing code, in running and analysing the experiments, and in drafting and editing the text and figures. The author reviewed and verified the code, data, and text and confirmed the final upload. A polished version will be updated soon. The Supplementary Information will be provided with the next version.

\subsection*{Acknowledgements}
Y. W. acknowledges support from Hong Kong Polytechnic University (Presidential Young Scholars Scheme, Project No. P0065105)

\end{document}